# ASTROMETRIC DISCOVERY OF GJ 802b: IN THE BROWN DWARF OASIS?

Steven H. Pravdo
Jet Propulsion Laboratory, California Institute of Technology
306-431, 4800 Oak Grove Drive, Pasadena, CA 91109; spravdo@jpl.nasa.gov

Stuart B. Shaklan
Jet Propulsion Laboratory, California Institute of Technology
301-486, 4800 Oak Grove Drive, Pasadena, CA 91109, shaklan@huey.jpl.nasa.gov

&
James Lloyd
Department of Astronomy, Cornell University
610 Space Sciences Building, Ithaca, NY 14853-6801, jpl@astro.cornell.edu

## ABSTRACT

The Stellar Planet Survey is an ongoing astrometric search for giant planets and brown dwarfs around a sample of ~30 M-dwarfs. We have discovered several low-mass companions by measuring the motion of our target stars relative to their reference frames. The lowest mass discovery thus far is GJ 802b, a companion to the M5-dwarf GJ 802A. The orbital period is 3.14 ± 0.03 y, the system mass is 0.214 ± 0.045 $M_\odot$, and the semi-major axis is 1.28 ± 0.10 AU or 81 ± 6 *mas*. Imaging observations indicate that GJ 802b is likely to be a brown dwarf with the astrometrically determined mass 0.058 ± 0.021 $M_\odot$ (one sigma limits). The remaining uncertainty in the orbit is the eccentricity that is now loosely constrained. We discuss how the system age limits the mass and the prospects to further narrow the mass range when *e* is more precisely determined.

## 1. INTRODUCTION

The Stellar Planet Survey (STEPS) is an astrometric search for low mass companions to M-dwarfs. Astrometry provides the most direct measurements of total system and component masses in planetary and binary systems. Even so, the results are dependent upon the parallax and the mass of the primary, the latter often derived through the mass-luminosity relationship (MLR). Independent knowledge of the primary mass avoids the degeneracy in the astrometric model between the total mass and the fractional mass—as the total mass increases the fractional mass can decrease to create the same astrometric signal.

Still, the uncertainties in the parallax and the MLR for main-sequence dwarfs are typically far less than the *additional* uncertainties that arise in model calculations of brown dwarf (BD) and planetary masses from physical principles alone. When we measure their masses, we test and assist the development of the models based upon parameters such as age and metallicity. Determining an accurate mass thus deepens our understanding of the fundamental physics of stars and substellar objects. Another direct benefit is to advance our knowledge of the mass-luminosity relationships (MLRs) for such objects to guide further research. At present there are no extant observational MLRs for brown dwarfs (BDs) and the MLR for stars at the bottom of the main sequence is based upon only 10 objects (Henry et al 1999). We have made several mass measurements of companions to M-dwarfs with the Stellar Planet Survey (STEPS, Pravdo & Shaklan 2003, Pravdo, Shaklan, Henry, & Benedict 2004--PSHB, Pravdo, Shaklan, Lloyd, & Benedict 2005--PSLB). In each case the combination of astrometry and imaging resulted in conclusions about the masses of the components that could not have been reached by either technique alone.

Hundreds of low mass objects have been discovered and studied since the advent of sensitive infrared programs such as 2MASS. These objects comprise a significant fraction of the stellar population and mass (e.g. Burgasser 2004). Classification systems for late M (Kirkpatrick, Henry, & McCarthy 1991, Kirkpatrick, Henry, & Simons 1995), L (Kirkpatrick et al. 1999), and T-dwarfs (Burgasser et al. 2002) have made remarkable progress. Spectral and mass models have followed (e.g. Chabrier et al. 2000, Burrows et al. 2000, Baraffe et al. 2003). The discoveries of BDs in systems (Reid et al. 2001, Freed, Close, & Siegler 2003, Close et al. 2003, Burgasser et al. 2003, Siegler et al. 2003, Golimowski et al. 2004, McCaughrean et al. 2004) have led to more robust estimates of masses, but there are currently few dynamical mass measurements of L and T dwarfs (Buoy et al. 2004, Zapetoro Osorio et al. 2004, Close et al. 2005).

## 2. OBSERVATIONS AND RESULTS

### 2.1 Astrometry

GJ 802 (=LHS 498, G 231-13, Wolf 1084) is an M dwarf with the properties listed in Table 1.We observed GJ 802 from 1998-2004 with the STEPS instrument mounted at the Cassegrain focus of the Palomar 200” (5-m) telescope. The first observation was July 3.4, 1998 = JD 2450997.9. PSHB and PSLB give more detailed descriptions of the instrument and data analysis.

Table 2 shows the results of our measurements of parallax and proper motion. Our parallax is measured relative to the in-frame reference and should be corrected for the reference frame’s finite distance. It is consistent with the currently accepted value (Table 1), with or without the addition of the 2 *mas* correction from relative to absolute parallax for average fields at this galactic latitude and apparent magnitude (van Altena, Lee, & Hoffleit 1995). Our proper motion values are also consistent with prior results at slightly more than 1 sigma where the error bars on the prior results are estimated from the variation among past observers (Luyten 1979, Harrington & Dahn 1980, Bakos, Sahu, & Nemeth 2002). In principle, our proper motions should be corrected for the average proper motion of the field, but this is a small effect and does not contribute to errors in the analysis below.

GJ 802 has a periodic astrometric signal after subtraction of parallax and proper motion from the total motion, indicating the presence of a companion, GJ 802b. Fig. 1 shows the astrometric data superimposed on an orbit with an acceptable fit. Our error estimates comprise the uncertainty due to the Poisson statistics of the image photon counts added in quadrature to 1.0 *mas* systematic errors. We determine the 1-sigma confidence limits in our observed parameters via the method described in Lampton, Margon, & Bowyer (1976) for multi-parameter estimation.

## 2.2 Adaptive Optics Imaging

We use an imaging observation to further constrain the system. We performed *H*-band adaptive optics (AO) observations with the Palomar 200" (5-m) system (Troy et al. 2000) on June 6, 2004 and Sept. 2, 2004 UT. Fig. 2 shows the resulting image of GJ 802 from Sept. The conditions were excellent on both nights with sub-arcsec seeing. We also show a comparison image of GJ 1210 that reveals its binary nature (PSLB) obtained during the June run. The failure to detect GJ 802b with AO rules out a companion within 3.25 *H*-magnitude (5%) of the primary. The components were separated by ~100 *mas* during the AO observation (Fig. 1), i.e, capable of being resolved (Fig. 2).

# 3. DISCUSSION

## 3.1 The Primary

GJ 802A is a field M dwarf 15.9 pc from the sun. It is active, classified as dM5e, with both Hα (Reid, Hawley, & Gizis 1995) and X-ray emission (Hünsch et al 1999). Its (*U,V,W*) space velocity is consistent with the local volume-complete sample of M dwarfs studied by Reid, Hawley, & Gizis (1995) although it is slightly farther away. Independent knowledge of the age and mass of the primary would be helpful in further constraining the properties of this system. An estimate of the age based upon the $V\text{-}I_C$ color at which such stars become active is ~6 Gy (Hawley et al. 1999). If we assume that all the light comes from the primary, its mass inferred from the *V* MLR (Henry et al. 1999, eq. 7) is consistent at the one sigma level with that inferred from the *H* MLR (Henry & McCarthy 1993, eq. 3a): $M_{pri}(V) = 0.150\ (+0.022,-0.029)\ M_\odot$ and $M_{pri}\ (H) = 0.174\ (+0.023,-0.020)\ M_\odot$. If the mass of the secondary takes its maximum acceptable value of 0.08 Mʋ (see following section), then $M_{pri}(V)$ is reduced by only 0.001 and $M_{pri}\ (H)$ is reduced by only 0.003 $M_\odot$ We therefore adopt $M_{pri} = 0.16 \pm 0.03\ M_\odot$ (cf. Close et al. 2005 for another view of the MLR accuracies).

Bonfils et al. (2005) give metallicity distributions of M-dwarf stars in the solar neighborhood. If we use his equation (1) to determine the metallicity of GJ 802A we find [Fe/H] = 0.025 if all the *V*- and *K*-band light came from the primary, and [Fe/H] = -0.042 in the other extreme, if the light were evenly divided between the components. However, if we apply his MLR (equation (2)) to this system, we find that it predicts a mass on the low end of our range for the primary, $M_{pri} = 0.13$, if it contains all the light. To get a mass more consistent with other MLRs would require [Fe/H] = 0.25, the upper limit of his range of validity. We conclude that there is no evidence for non-solar metallicity but a question remains about the consistency of the MLRs.

## 3.2 The System

Our astrometric measurements yield the orbital parameters subject to two ambiguities. First, the scale of the system is not uniquely determined because we do not resolve the components. Thus, for a given period, the data admit a range of values for the semi-major axis, *a*, and total mass, $M_{tot} = M_{pri} + M_{sec}$. This is shown in Figure 3. The open triangles show only the acceptable fits to the data after ~11000 Monte Carlo trials. The Y axis of Figure 3 is our observed parameter, $(f\text{-}\beta) = \alpha/a$, where, $f = M_{sec}/M_{tot}$, the fractional mass, β is the fractional light, and $\alpha/a$ is the ratio of the photocentric to Keplerian orbits (e.g. PSHB). Second, the value of ($f$-β) can be the same in two very different physical situations. If the secondary is small in mass compared to the primary, we have: $f \ll 1$, $\beta \sim 0$, and $(f-\beta) \sim f \ll 1$. Conversely, if the secondary is close in mass to the primary, we have: $f \sim 0.5$, $\beta \sim 0.5$, but also, $(f-\beta) \ll 1$.

Fortunately we can use other information to resolve these ambiguities. The total mass of the system is bounded by the mass limits on the primary based upon its spectrophotometry. Additionally, the values for $f$ and β are related by a MLR. A current observational $V$-band MLR (Henry et al. 1999) is based upon stars with masses from 0.074-0.178 $M_\odot$. Since the $V$ light contribution is negligible for masses « 0.08 $M_\odot$, we create a β = 0 region that allows us to extend the curve for the Henry et al. MLR into the BD range (solid line in Fig. 3). We also illustrate a 5-Gy model from Baraffe et al. (2003) that already extends throughout the BD realm (filled circles in Fig. 3). The MLRs agree well with each other below the peak of ($f$-β). The fits to the STEPS data that overlie the MLRs represent the orbital models consistent with all the currently known information.

The fact that the MLRs in Fig. 3 have two $M_{tot}$ values for each ($f$-β) illustrates the second ambiguity mentioned above. These are the high ($f \sim 0.5$) and low ($f \ll 1$) mass branches. However, our AO observations eliminate the high mass branch. The $M_H$ of the GJ 802 composite source is 8.05 based upon the 2MASS measurement and the parallax (Table 1). $M_H$ of GJ 802b is then > 11.36 based upon our AO observation. This value is ~3 times fainter than that for the lowest applicable mass of the $H$-band MLR for late M dwarfs (Henry & McCarthy 1993), and implies a secondary mass, $M_{sec} < 0.08$ $M_\odot$. The implied $V$ luminosity of such an object compared with the total $V$ luminosity of GJ 802 results in $\beta < 0.024$. This also places it in the ascending portion of the ($f$-β) function and rules out the high mass branch. Fig. 4 shows the GJ 802 $H$-band secondary-to-primary ratios for the Baraffe BD models. Values greater than 0.05 for the $H$-band ratio are ruled out by our AO observations. Therefore, for the ages of BDs shown, mass values to the right of where the curves intercept 0.05 are ruled out. The upper limits are 0.078, 0.073, and 0.060 $M_\odot$, for 5, 1, and 0.5 Gy, respectively. The 5 Gy upper limit is probably applicable based upon other indicators of the system age (see §3.1). Another version of the MLR (Delfosse et al. 2000) is applicable only in the high-mass branch shown in Fig. 3, and results in estimates ~0.025 $M_\odot$ higher that the other models shown.

Table 2 lists the orbital parameters. The major remaining uncertainty in the orbit is the eccentricity $e$. The mass of GJ 802b is dependent on $e$ as shown in Fig. 5, and is now constrained to be 0.057 ± 0.021 $M_\odot$ (37-82 Jupiter masses, $M_J$).

### 3.3 GJ 802b

There are a number of observational possibilities to further constrain the mass of GJ 802b. Continuing STEPS astrometry will succeed if we are able to obtain observations at a critical phase to distinguish among different values of $e$. The current uncertainty in $e$ is due to an unfavorable temporal beating between the observational opportunities and the period. Even limiting the eccentricity to values less than 0.5 will reduce the mass upper limit to 0.05 $M_\odot$ (52 $M_J$). Additionally, a Hubble Space Telescope/NICMOS imaging observation with its high spatial resolution and sensitivity in the $JHK$ region can not only measure the separation and position angle of the system but also the flux ratio for many

BD models (e.g. PSHB). Fig. 6 illustrates the relative spectra in this band normalized to a distance of 10 pc for an M dwarf similar to GJ 802A, a late T dwarf, and models for 40 and 60 $M_J$ BDs at age 5 Gy. Primary-to-secondary *JHK* flux ratios will be in the 30-200 range for different BD spectral types.

BDs are both X-ray and Hα emitters (e.g. Tsuboi et al. 2000). GJ 802b might be detectable in Hα at HST spatial resolution if the emission were as large as the 1-10 Å equivalent width. measured for other BDs, and at a contrast ratio of ~100. The X-ray emission is not yet separately measurable because current instruments are limited to ~1" spatial resolution. However, another factor of ~10 in spatial resolution and the companion would be detectable if it reached, for example, the peak of the flaring X-ray emission from the 0.5-Gy BD, LP 944-20, ~$10^{26}$ ergs $s^{-1}$, or ~1% of the total GJ 802 emission (Hünsch et al 1999).

The gains for more accurately measuring the mass of GJ 802b are twofold. First it will place a point on the BD MLR and continue the determination of that useful research tool. Second it offers the possibility of acquiring a spectrum to accompany the accurate mass measurement that will guide further model development.

### 3.4 The Brown Dwarf Oasis?

The BD desert may prove to be a mirage when one knows where to search. STEPS is unique in its target set of nearby M stars and its ability to astrometrically probe close to the primary at the secondary mass limits described above. We chose all of our targets (excepting a known control) because they were single stars, according to the state of the science in 1997. We have now detected 5 companions to 24 targets that have sufficient data. Three of the companions are late M stars, two are in or near the BD range (GJ 802b and GJ 164B), and for the others, the existences and masses of potential companions are still pending. Although the STEPS numbers are small, even now the percentages are inconsistent with the presumed BD desert—e.g. < 1% of solar-type stars have a brown dwarf within 5 AU (Marcy, Cochran, & Mayor 2000).

## Table 1. GJ 802 Known Properties

| | |
|---|---|
| **RA (2000)[a]** | 20 43 19.41 |
| **Dec (2000)[a]** | +55 20 52.0 |
| ***V*[b]** | 14.69 |
| ***J*[c]** | 9.563 ± 0.023 |
| ***H*[c]** | 9.058 ± 0.019 |
| ***K*[c]** | 8.753 ± 0.013 |
| **Type** | dM5e |
| **Parallax[d] (*mas*)** | 63 ± 5.5 |
| **Proper Motion[e] (*mas* $y^{-1}$)** | 1915 ± 13 |
| **Position Angle[e] (deg)** | 27.6 ± 0.6 |

[a]Bakos, Sahu, & Nemeth (2002), [b]Weis 1988, [c]2MASS,[d]van Altena, Lee & Hoffleit 1995, [e]Luyten 1979

## Table 2. STEPS Astrometric Measurements[a] of GJ 802

| | |
|---|---|
| **Relative Parallax (*mas*)** | 61 ± 2 |
| **Proper Motion (*mas* $y^{-1}$)** | 1933 ± 1 |
| **Position Angle (deg)** | 27.0 ± 0.1 |
| **Period (y)** | 3.14 ± 0.03 |
| **Total Mass ($M_\odot$)** | 0.215 ± 0.045 |
| **Semi-Major Axis (AU)** | 1.28 ± 0.10 |
| **Eccentricity, *e*** | 0.56 ± 0.30 |
| **Inclination (deg)** | 80.5 ± 1.5 |
| **Lon. Asc. Node[b] (deg)** | 17.5 ± 3.5 |
| **Primary Mass, $M_{pri}$ ($M_\odot$)** | 0.160 ± 0.03 |
| **Secondary Mass, $M_{sec}$ ($M_\odot$)** | 0.057 ± 0.021 |

[a]epoch and argument of the periapse are not meaningfully constrained
[b]or + 180º because of into or out of plane ambiguity

### ACKNOWLEDGMENTS

The research described in this paper was performed in part by the Jet Propulsion Laboratory, California Institute of Technology, under contract with the National Aeronautics and Space Administration. We thank B. Oppenheimer and A. Gould for useful discussions. We performed observations at Caltech's Palomar Observatory and acknowledge the assistance of the staff. This research has made use of the NASA/IPAC Infrared Science Archive, which is operated by the Jet Propulsion Laboratory, California Institute of Technology, under contract with the National Aeronautics and Space Administration. This research has made use of the SIMBAD database, operated at CDS, Strasbourg, France, and of NASA's Astrophysics Data System Abstract Service. This publication makes use of data products from the Two Micron All Sky Survey, which is a joint project of the University of Massachusetts and the Infrared Processing and Analysis Center/California Institute of Technology, funded by the National Aeronautics and Space Administration and the National Science Foundation.

## FIGURE CAPTIONS

1. The STEPS data (points) are superimposed on a model of the Keplerian circular orbit (solid red lines). The RA and Decl. dimensions vs. time are shown separately. The 1-sigma error bars on the points are our photocentric measurement errors multiplied by the ratio of the Keplerian to the photocentric orbit. The position of the AO observation is also shown (green filled circle).

2. Palomar 200" (5-m) AO image of GJ 802 (left panel) and GJ 1210 (right panel). The scale is the same for both images.

3. The points show the results of ~11,000 Monte Carlo trials for the GJ 802 orbit (open triangles). We plot ($f$-β) vs. $M_{tot}$ for all models falling within the one-sigma confidence limits. Superimposed on the data are the composite MLR curve in the *V*-band based upon observations (Henry et al. 1999, solid line) and the MLR points (filled circles) from the model of Baraffe et al. (2003).

4. The *H*-band ratios for different system ages based upon the models of Baraffe et al. (2003).

5. The secondary mass, $M_{sec}$, as a function of eccentricity, $e$, for models with 1,2, and 3-σ confidence limits.

6. Comparison of the near-IR spectra normalized to 10 pc of an M dwarf (M5.5), a late T dwarf (T6), and two models for age 5-Gy BDs with masses of 40 and 60 $M_J$. The observed spectra are from McLean et al. (2003) and the models from Burrows et al. (2002) and Burrows, Sudarsky, & Lunine (2003)

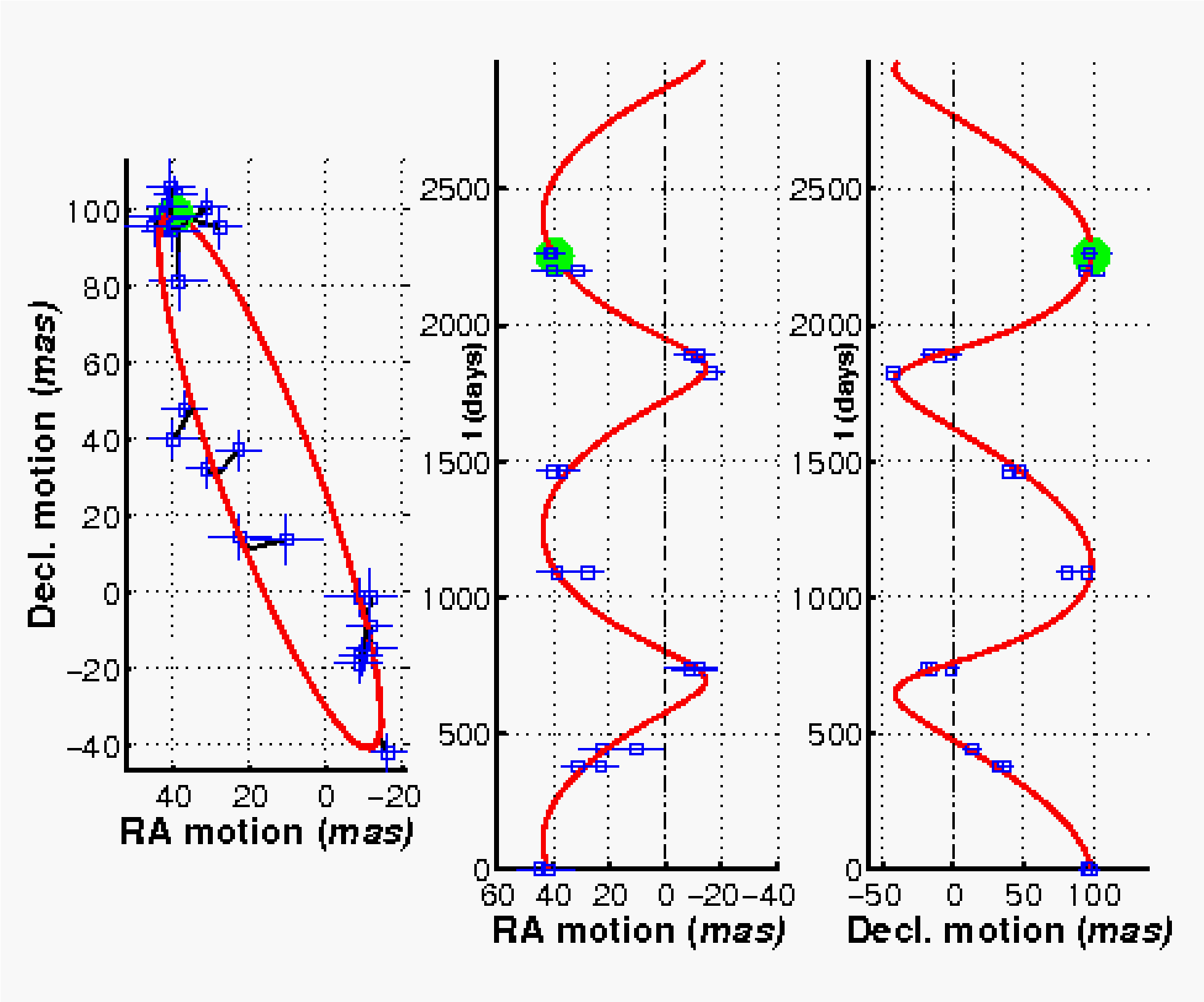

**Figure 1**

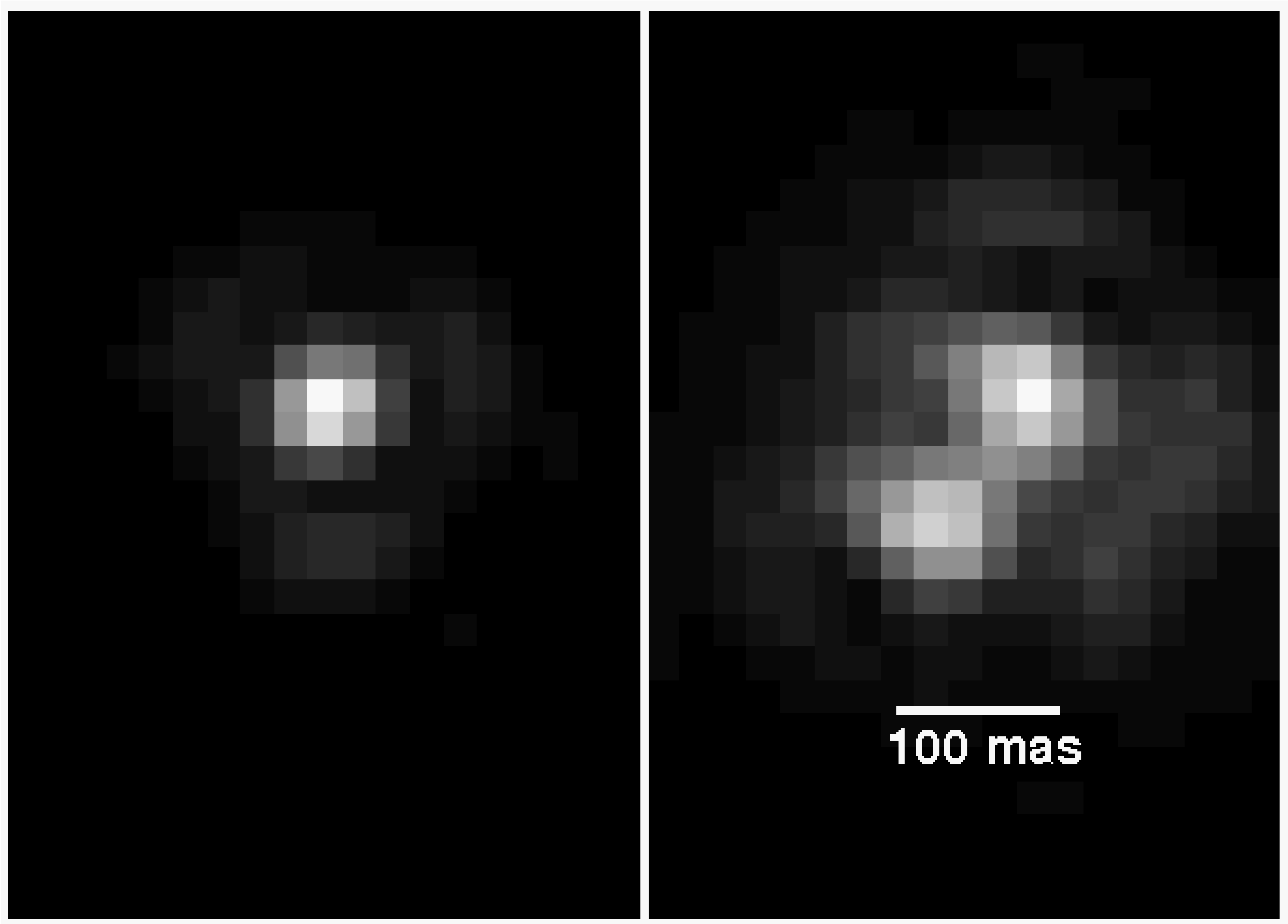

Figure 2

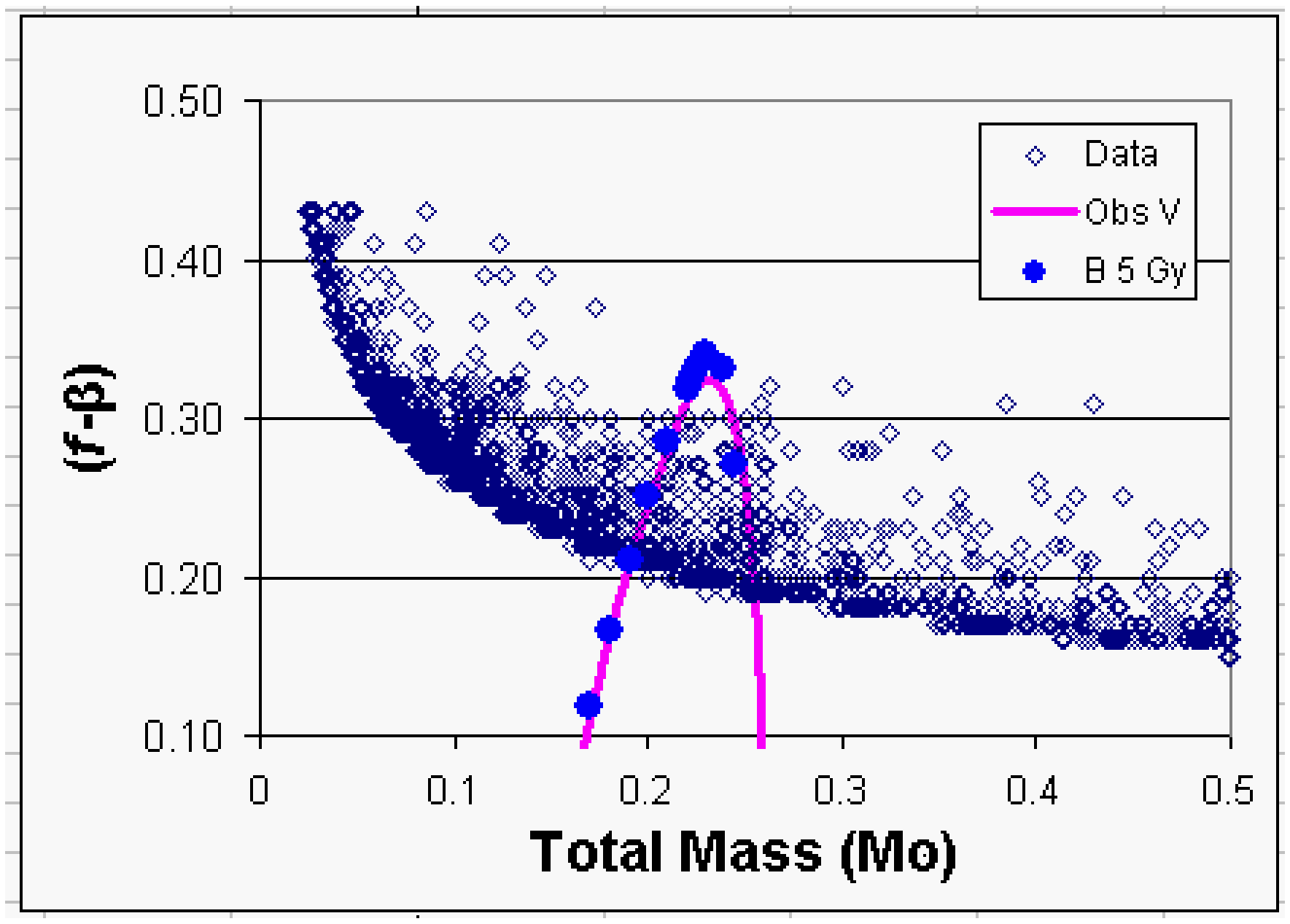

Figure 3

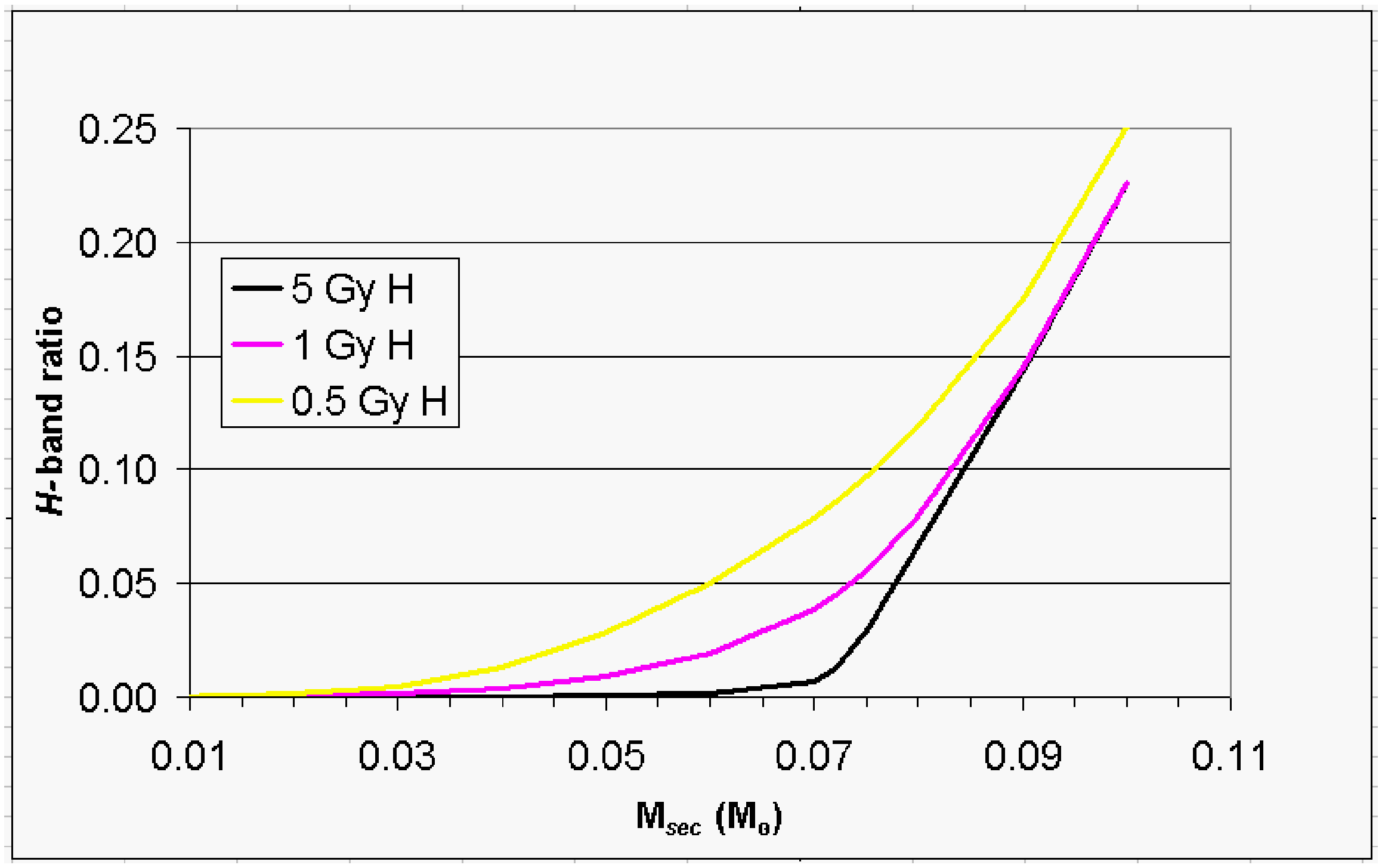

**Figure 4**

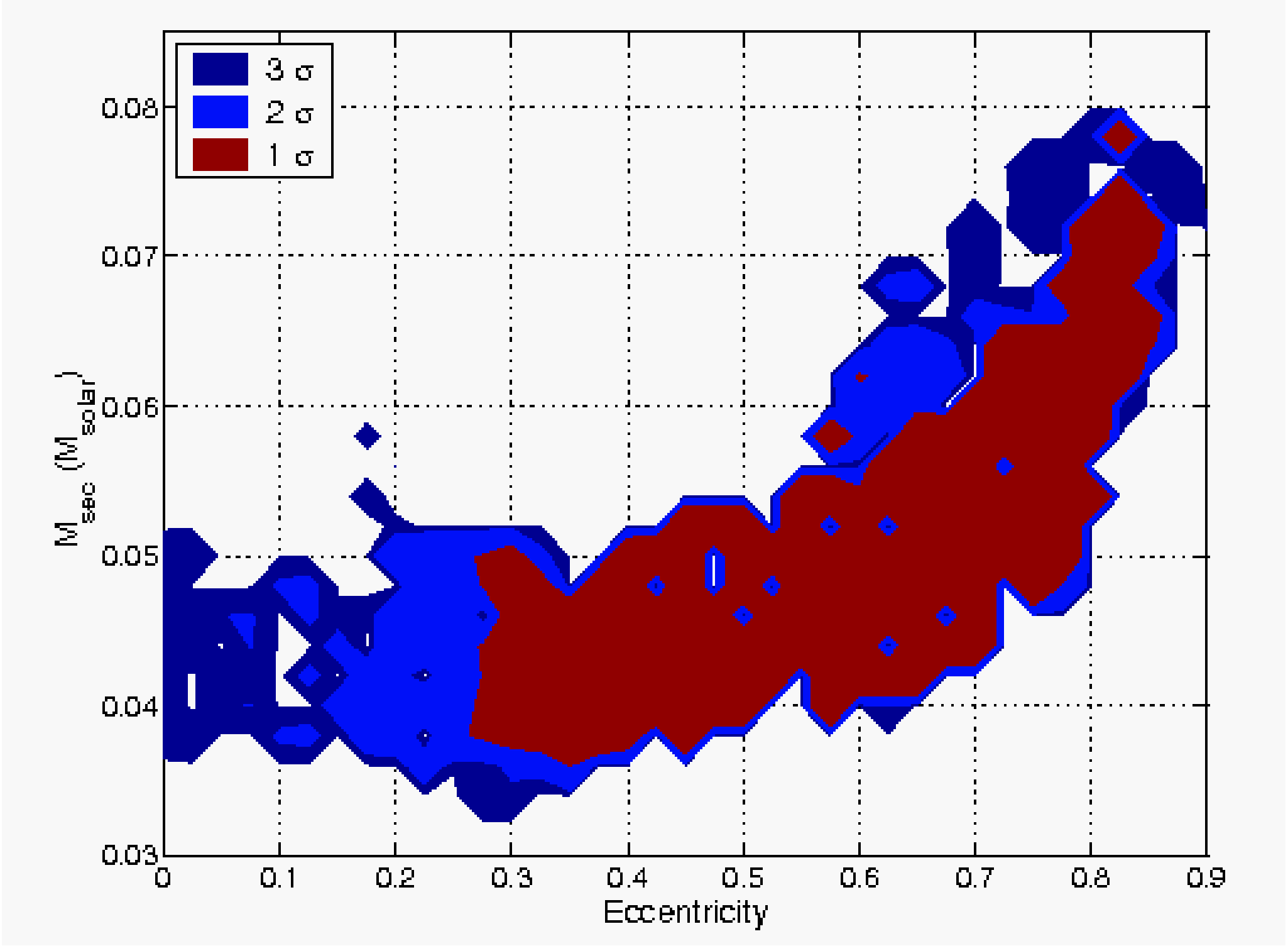

**Figure 5**

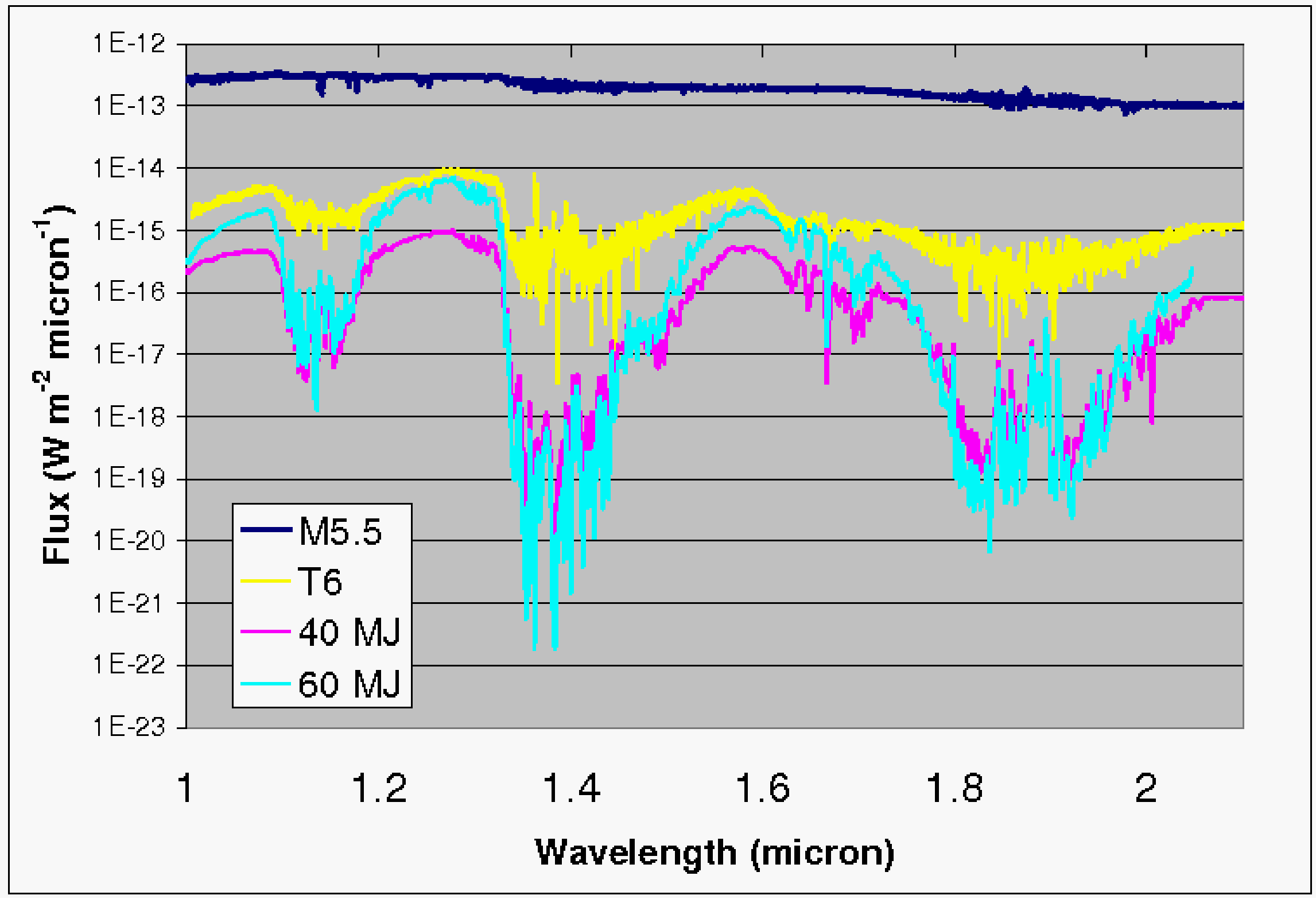

**Figure 6**